\documentstyle[12pt]{article}

\newfont{\Mb}{msbm10}                    
\newcommand{\R}{\mbox{\Mb\symbol{82}}}   
\newtheorem{theorem}{Theorem}            
\newcommand{\sds}{\mbox{$\subset \!\!\!\!\!\!+$}}

\begin{document}

\renewcommand{\baselinestretch}{1.07}

\title{\Large \bf G\"odel-type Spacetimes in Induced Matter\\
                   Gravity Theory \\}
\author{
H.L. Carrion\thanks{ {\sc e-mail:}  lenyj@cbpf.br }, \ \  
M.J. Rebou\c{c}as\thanks{{\sc e-mail:} reboucas@cbpf.br} \ \ 
and  \  
A.F.F. Teixeira\thanks{{\sc e-mail:} teixeira@cbpf.br} \\ 
\\  
Centro Brasileiro de Pesquisas F\'\i sicas\\
Departamento de Relatividade e Part\'\i culas \\
Rua Dr.\ Xavier Sigaud 150 \\
22290-180 Rio de Janeiro -- RJ, Brazil \\ \\
       }

\date{\today}  

\maketitle

\begin{abstract}  
A f\/ive-dimensional (5D) generalized G\"odel-type 
manifolds are examined in the light of the equivalence problem 
techniques, as formulated by Cartan.
The necessary and suf\/f\/icient conditions for local homogeneity 
of these 5D manifolds are derived. 
The local equivalence of these homogeneous Riemannian manifolds 
is studied. It is found that they are characterized by three essential 
parameters $k$, $m^2$ and $\omega\,$: identical triads $(k, m^2, \omega)$ 
correspond to locally equivalent 5D manifolds. An irreducible
set of isometrically nonequivalent 5D locally homogeneous Riemannian 
generalized G\"odel-type metrics are exhibited. A classif\/ication of these 
manifolds based on the essential parameters is presented, 
and the Killing vector f\/ields as well as the corresponding Lie algebra 
of each class are determined. It is shown that the generalized
G\"odel-type 5D manifolds admit maximal group of isometry $G_r$
with $r=7$, $r=9$ or $r=15$ depending on the essential parameters
$k$, $m^2$ and $\omega\,$. The breakdown of causality in
all these classes of homogeneous G\"odel-type manifolds are also 
examined. It is found that in three out of the six irreducible
classes the causality can be violated. The unique generalized
G\"odel-type solution of the induced matter (IM) f\/ield equations
is found. 
The question as to whether the induced matter version of general
relativity is an ef\/fective therapy for these type of causal 
anomalies of general relativity is also discussed in connection 
with a recent work by Romero, Tavakol and Zalaletdinov.
\end{abstract}

\newpage

\section{Introduction}   \label{intro}
\setcounter{equation}{0}

The f\/ield equations of the general relativity theory, which
in the usual notation are written in the form 
\begin{equation}
G_{\alpha \beta} = \kappa \,\; T_{\alpha \beta}\;, \label{ein} 
\end{equation}
relate the geometry of the spacetime to its source. The general 
relativity theory, however, does not prescribe 
the various forms of matter, and takes over the energy-momentum 
tensor $\,T_{\alpha \beta}\,$ from other branches of physics. 
In this sense, general relativity (GR) is not a closed theory. 
The separation between the gravitational f\/ield and its source 
has been often considered as one undesirable feature of GR%
~\cite{Einstein56}~--~\cite{Salam80}.

Recently, Wesson and co-workers~\cite{Wesson90,WessonLeon92a}
have introduced a new approach to GR, in which 
the matter and its role in the determination of the spacetime 
geometry is given from a purely f\/ive-dimensional geometrical 
point of view. In their f\/ive-dimensional (5D) version of general 
relativity the f\/ield equations are given by 
\begin{equation} \label{5DfeqsG}
\widehat{G}_{AB} = 0 \;.
\end{equation}
Henceforth, the f\/ive-dimensional geometrical objects are denoted by 
overhats and Latin letters are 5D indices and run from $0$ to $4$.
In this new approach to GR the 5D vacuum f\/ield equations%
~(\ref{5DfeqsG}) give rise to both curvature and matter in 4D. 
Indeed, it can be shown~\cite{WessonLeon92a} that it is always 
possible to rewrite the f\/ifteen f\/ield equations~(\ref{5DfeqsG}) 
as a set of equations such that ten of which are precisely Einstein's 
f\/ield equations~(\ref{ein}) in 4D with an {\em induced\/} 
energy-momentum 
\begin{eqnarray} \label{Tinduced}
\kappa \; T_{\alpha\beta} & = & \frac{\phi_{\alpha\,;\:\beta}}{\phi} - 
\frac{\varepsilon}{2\,\phi^2} 
   \left\{ \frac{\phi^{*} \, g^{*}_{\alpha\beta}}{\phi}
   - g^{**}_{\alpha\beta} + g^{\gamma\delta} \, 
   g^{*}_{\alpha\gamma} \,  g^{*}_{\beta\delta} 
-  \frac{g^{\gamma\delta} \, 
   g^{*}_{\gamma\delta} \, g^{*}_{\alpha\beta}}{2}  
   \right. \nonumber \\
              &   &
+ \; \left.
\frac{g^{}_{\alpha\beta}}{4^{} {}_{} } \, \left[\,g^{*}{}^{\gamma\delta} \, 
 g^{*}_{\gamma\delta} + (g^{\gamma\delta_{}} \,g^{*}_{\gamma\delta})^2
       \, \right]  \,\right\} \;,
\end{eqnarray} 
where the Greek letters denote 4D indices and run from $0$ to $3$, 
$g_{44} \equiv \varepsilon\, \phi^2$ with $\varepsilon=\pm 1 $,
$\phi_\alpha \equiv \partial \phi / \partial x^\alpha$, a star
denotes $\partial / \partial x^4$, and a semicolon  denotes the usual
4D covariant derivative. Obviously,
the remaining f\/ive equations (a wave equation and four conservation 
laws) are automatically satisf\/ied by any solution of the 5D 
vacuum equations~(\ref{5DfeqsG}).
Thus, not only the matter  but also its role in the 
determination of the geometry of the 4D spacetime can be considered 
to have a f\/ive-dimensional geometrical origin. This approach 
unif\/ies the gravitational f\/ield with its source (not just 
with a particular f\/ield) within a purely 5D geometrical framework.
This 5D version of general relativity is often referred to as 
induced matter gravity theory (IM gravity theory, for short). 
The IM theory has become a focus of a recent research f\/ield%
~\cite{Overduin97}.
The basic features of the theory have been explored by Wesson and 
others~\cite{Leon93}~--~\cite{Wesson96a}, whereas the implications 
for cosmology and astrophysics have been investigated by a number 
of researchers~\cite{Wesson92b}~--~\cite{Liu96b}. 
For a fairly updated list of references on IM  gravity theory 
and related issues we refer the reader to Overduin and 
Wesson~\cite{Overduin97}.

In general relativity, the causal structure of 4D spacetime
has locally the same qualitative nature as the f\/lat spacetime
of special relativity --- causality holds locally. The global
question, however, is left open and signif\/icant dif\/ferences can 
occur. On large scale, the violation of causality is not excluded. 
Actually, it has long been known that there are solutions to 
the Einstein f\/ield equations which possess causal anomalies in 
the form of closed timelike curves. The famous solution found 
by G\"odel~\cite{Godel49} in 1949 might not be the f\/irst but
it certainly is the best known example of a cosmological
model which makes it apparent that general relativity,
as it is normally formulated, does not exclude the existence of 
closed timelike world lines, despite its Lorentzian character 
which leads to the local validity of the causality principle.
Owing to its striking properties G\"odel's model
has a well-recognized importance and has to a certain extent 
motivated the investigations on rotating cosmological
G\"odel-type models and on causal anomalies in the 
framework of general relativity~\cite{Som68}~--~\cite{Krasinski98} 
and other theories of gravitation~\cite{Vaidya84}~--%
~\cite{FonsecaReboucas98}.

Two recent articles have been concerned with {\em five-dimensional\/} 
G\"odel-type spacetimes. Firstly in Ref.~\cite{ReboucasTeixeira98a}
the main geometrical properties of f\/ive-dimensional 
Riemannian manifolds endowed with a 5D counterpart of the 4D 
G\"odel-type metric of general relativity were investigated.
Among several results,  an irreducible set of isometrically 
nonequivalent 5D homogeneous (locally)  G\"odel-type 
metrics were exhibited. Therein it was also shown that, 
apart from the degenerated G\"odel-type metric, in all 
classes of homogeneous G\"odel-type geometries there is 
breakdown of causality. As no use of any particular f\/ield 
equations was made in this f\/irst paper, its results hold for 
any 5D G\"odel-type manifolds regardless of the underlying 
5D Kaluza-Klein gravity theory.  In the second article%
~\cite{ReboucasTeixeira98b} the classes of 5D G\"odel-type 
spacetimes discussed in~\cite{ReboucasTeixeira98a}
were investigated from a more physical viewpoint. 
Particularly, it was examined the question as to 
whether the induced matter theory of gravitation permits the
family of noncausal solutions of G\"odel-type metrics studied 
in~\cite{ReboucasTeixeira98a}. It was shown that the 
IM gravity excludes this class of 5D G\"odel-type 
non-causal geometries as solution to its f\/ield equations. 

In both articles~\cite{ReboucasTeixeira98a,ReboucasTeixeira98b} 
the 5D G\"odel-type family of metrics discussed is the simplest 
5D class of geometries for which the section $u = \mbox{const}$ 
($u$ is the extra coordinate) is the 4D G\"odel-type metric of 
general relativity.  Actually the 5D G\"odel-type line element
of both papers does not depend on the f\/ifth coordinate
$u$, and therefore as regards to the IM theory a radiation-like 
equation of state is an underlying assumption of both articles.
However, it is well know~\cite{Overduin97} that the dependence 
of the 5D metric on the extra coordinate is necessary to ensure 
that the 5D IM theory permits the induction of matter of a very 
general type in 4D.

In this work, on the one hand, we shall examine the main 
geometrical properties of a class of {\em generalized\/} 
G\"odel-type geometries in which the 5D metric depends on the 
f\/ifth coordinate, generalizing therefore the results found in 
Ref.\ \cite{ReboucasTeixeira98a}. On the other hand, 
we shall also investigate the question as to whether 
the induced matter gravity theory, as formulated 
by Wesson and co-workers~\cite{Wesson90,WessonLeon92a},
admits these generalized G\"odel-type metrics as solutions 
to its f\/ield equations, thus also extending the
investigations of Ref.\  \cite{ReboucasTeixeira98b}.

The outline of this article is as follows.
In the next section we present a summary of some important
prerequisites for Section~3, where using the equivalence 
problem techniques as formulated by Cartan~\cite{Cartan}
we derive the necessary and suf\/f\/icient conditions for local
homogeneity of this class of 5D generalized  
G\"odel-type manifolds. In Section~3 we also exhibit an 
irreducible set of isometrically nonequivalent homogeneous 
generalized G\"odel-type metrics. In Section~4 we discuss  
the integration of the Killing equations and present the Killing 
vector f\/ields as well as the corresponding Lie algebra for 
all homogeneous generalized G\"odel-type metrics.
In the last section we examine whether the IM f\/ield 
equations permit solutions of this generalized 
G\"odel-type class of geometries. The unique solution of
this type is found therein. 
The question as to whether the IM version of general 
relativity rules out the existence of closed
timelike curves of G\"odel type is also discussed 
(Section~5) in connection with a recent paper by 
Romero {\em et al.\/}~\cite{Romero96}.

\vspace{3mm}
\section{Prerequisites}
\label{prereq}
\setcounter{equation}{0}

The arbitrariness in the choice of coordinates in the metric
theories of gravitation gives rise to the problem of deciding
whether or not two manifolds whose metrics $g$ and $\tilde{g}$ 
are given explicitly in terms of coordinates, viz.,
\begin{equation}
ds^2         =        g_{\mu \nu} \,dx^\mu \, dx^\nu \qquad \;
\mbox{and} \qquad \;  
d\tilde{s}^2 = \tilde{g}_{\mu \nu}\,d\tilde{x}^\mu\,d\tilde{x}^\nu\:,
\end{equation}
are locally isometric. This is the so-called equivalence problem (see 
Cartan~\cite{Cartan} for the local equivalence of $n$-dimensional 
Riemannian manifolds, Karlhede~\cite{Karlhede80} and
MacCallum~\cite{MacCallumSkea94} for the special case $n=4$ of 
general relativity).

\begin{sloppypar}
The Cartan solution~\cite{Cartan} to the equivalence problem 
for Riemannian manifolds can be summarized as follows. Two 
$n$-dimensional Lorentzian Riemannian manifolds ${\cal M}_n$ and 
$\widetilde{\cal M}_n$ are locally equivalent if there exist coordinate 
and generalized $n$-dimensional Lorentz transformations such that 
the following {\em algebraic} equations relating the frame components 
of the curvature tensor and their covariant derivatives: 
\end{sloppypar}
\parbox{14cm}{\begin{eqnarray*} 
R^{A}_{\ BCD} & = &  \widetilde{R}^{A}_{\ BCD}\:, \nonumber \\
R^{A}_{\ BCD;M_1} & = & \widetilde{R}^{A}_{\ BCD;M_1}\:, \nonumber  \\
R^{A}_{\ BCD;M_1 M_2} &=&\widetilde{R}^{A}_{\ BCD;M_1 M_2}\:,\nonumber  \\
                  & \vdots &   \nonumber \\
R^{A}_{\ BCD;M_1\ldots M_{p+1}} & = & \widetilde{R}^{A}_{\ BCD;M_1
                                         \ldots M_{p+1}} \nonumber \\  
            \end{eqnarray*}}  \hfill
\parbox{1cm}{\begin{eqnarray}   \label{eqvcond}  \end{eqnarray}}
are compatible as {\em algebraic} equations in  
$\left( x^{\mu}, \xi^{A} \right)$. Here and in what follows we use 
a semicolon to denote covariant derivatives. Note that $x^{\mu}$ 
are coordinates on the manifold ${\cal M}_n$ while $ \xi^{A}$
parametrize the group of allowed frame transformations 
[$n$-dimensional generalized Lorentz group usually denoted%
~\cite{HawkingEllis73} by $O(n-1, 1)\,$]. Reciprocally, equations 
(\ref{eqvcond}) imply local equivalence between the $n$-dimensional 
manifolds ${\cal M}_n$ and $\widetilde{\cal M}_n$.

In practice, a f\/ixed frame is chosen to perform the calculations so 
that only coordinates appear in the components of the curvature tensor,
i.e. there is no explicit dependence on the parameters $\xi^{A}$ of the 
generalized Lorentz group.

Another important practical point to be considered, once one wishes 
to test the local equivalence of two Riemannian manifolds, is that 
before attempting to solve eqs.\ (\ref{eqvcond}) one can extract and 
compare partial pieces of information at each step of dif\/ferentiation
as, for example, the number $\{t_{0},t_1, \dots ,t_{p}\}$ of 
functionally independent functions of the coordinates $x^\mu$
contained in the corresponding set
\begin{equation}   \label{CartanScl}
I_{p} = \{ R^{A}_{\ BCD} \,, \,R^{A}_{\ BCD;M_{1}} \,, \,
R^{A}_{\ BCD;M_1 M_2}\,,\,\ldots,\,R^{A}_{\ BCD;M_1 M_2\ldots M_{p}}\}\,,  
\end{equation} 
and the isotropy subgroup $\{H_{0}, H_1, \ldots ,H_{p}\}$ of the 
symmetry group $G_r$ under which the set corresponding $I_p$ is 
invariant. They must be the same for each step $q= 0, 1, \cdots ,p$ 
if the  manifolds are locally equivalent.

In practice it is also important to note that in calculating the 
curvature and its covariant derivatives, in a chosen frame, one 
can stop as soon as one reaches a step at which the $p^{th}$ 
derivatives (say) are algebraically expressible in terms of the 
previous ones, and the residual isotropy group  (residual frame 
freedom) at that step is the same isotropy group of the previous 
step, i.e.  $H_p = H_{(p-1)}$. In this case further dif\/ferentiation
will not yield any new piece of information.
Actually, if $H_p = H_{(p-1)}$ and, in a given frame, the $p^{th}$ 
derivative is expressible in terms of its predecessors, for any 
$q > p$ the $q^{th}$ derivatives can all be expressed in terms 
of the $0^{th}$, $1^{st}$, $\cdots$, $(p-1)^{th}$ derivatives%
~\cite{Cartan,MacCallumSkea94}.  

Since there are $t_p$  essential coordinates, in 5D 
clearly  $5-t_p$ are ignorable, so the isotropy group will 
have dimension  $s = \mbox{dim}\,( H_p )$, and the group of 
isometries of the metric will have dimension $r$ given by 
(see Cartan~\cite{Cartan})
\begin{equation}
r = s + 5 - t_p \:, \label{gdim}
\end{equation}
acting on an orbit with dimension
\begin{equation}
d = r - s = 5 - t_p \:.  \label{ddim}
\end{equation}

\vspace{3mm} 
\section{Homogeneity and Nonequivalent Metrics}
\label{homoconds}
\setcounter{equation}{0}

The line element of the f\/ive-dimensional {\em generalized\/}
G\"odel-type manifolds ${\cal M}_5$ we are concerned with 
is given by 
\begin{equation} \label{ds2a} 
d\hat{s}^{2} = dt^2 + 2\,H(x)\, dt\,dy - dx^2 - G(x)\,dy^2 - 
         \widetilde{F}^2(\tilde{u})\,(d\tilde{z}^2 + d\tilde{u}^2) \:,
\end{equation}
where $H(x)$, $G(x)$ and $\widetilde{F}(\tilde{u})$ are arbitrary 
real functions. 
By a suitable choice of coordinates the line element%
~(\ref{ds2a}) can be brought into the form
\begin{equation} \label{ds2} 
d\hat{s}^{2} = [\,dt + H(x)\,dy\,]^2 - dx^2 - D^2(x)\,dy^2 - 
                F^2(u)\,\,dz^2 - du^2 \:,
\end{equation}
where $D^2(x) = G + H^2$ and $u$ clearly is a new f\/ifth coordinate.

At an arbitrary point of ${\cal M}_5$ one can choose the following set 
of linearly independent one-forms $\widehat{\Theta}^A$: 
\begin{equation} \label{lorpen}
\widehat{\Theta}^{0} = dt + H(x)\,dy\:, \: \quad
\widehat{\Theta}^{1} = dx\:, \: \quad
\widehat{\Theta}^{2} = D(x)\,dy\:, \:\quad
\widehat{\Theta}^{3} = F(u)\, dz \:, \: \quad
\widehat{\Theta}^{4} = du \:,          
\end{equation}
such that the G\"odel-type line element (\ref{ds2}) can be 
written as
\begin{equation} \label{ds2f}
d\hat{s}^2 = \widehat{\eta}^{}_{AB} \: \widehat{\Theta}^A \,\, 
                   \widehat{\Theta}^B = 
(\widehat{\Theta}^0)^2 - (\widehat{\Theta}^1)^2 - 
(\widehat{\Theta}^2)^2 - (\widehat{\Theta}^3)^2 - 
(\widehat{\Theta}^4)^2\:. 
\end{equation}
Here and in what follows capital letters 
are 5D Lorentz frame indices and run from 0 to 4; they are raised 
and lowered with Lorentz matrices
$\widehat{\eta}^{AB} = \widehat{\eta}^{}_{AB} = 
                 \mbox{diag} (+1, -1, -1, -1, -1)$,
respectively. 

Using as input the one-forms (\ref{lorpen}) and the Lorentz frame
(\ref{ds2f}), the computer algebra package {\sc classi}%
~\cite{MacCallumSkea94,Aman87}, e.g., gives the following 
nonvanishing Lorentz frame components $\widehat{R}_{ABCD}$ of 
the curvature:
\begin{eqnarray}  
\widehat{R}_{0101} &=& \widehat{R}_{0202}=- \frac{1}{4} \, 
        \left( \frac{H'}{D}\, \right)^2\:,  \label{rie1st} \\  
\widehat{R}_{0112} & =& \frac{1}{2}\, \left(\frac{H'}{D}\,\right)' \:, 
                         \label{rie2nd} \\  
\widehat{R}_{1212} &=& \frac{D''}{D}-\frac{3}{4}\, 
            \left( \frac{H'}{D}\,\right)^2 \label{rie3rd}\:, \\
\widehat{R}_{3434} &=& \frac{\ddot{F}}{F} \label{rielast}  \;\,,
\end{eqnarray}
where the prime and the dot denote, respectively, derivative 
with respect to $x$ and $u$.

For 5D (local) homogeneity from eq.~(\ref{ddim}) one must have 
$t_q=0$ for $q=0, 1, \cdots\, p$, that is, the number of 
functionally independent functions of the coordinates $x^\mu$ in 
the set $I_p$ must be zero. Therefore, from eqs.%
~(\ref{rie1st})~--~(\ref{rielast}) we conclude that for 5D 
homogeneity it is necessary that
\begin{eqnarray}
\frac{H'}{D} &=&\mbox{const} \equiv -\,2\,\omega \label{cond1} \:, \\
\frac{D''}{D}&=&\mbox{const} \equiv m^2  \label{cond2} \:, \\
\frac{\ddot{F}}{F}&=&\mbox{const} \equiv k \:. \label{cond3}
\end{eqnarray}

The above necessary conditions are also suf\/f\/icient for 
5D local homogeneity. Indeed, under these conditions the 
nonvanishing frame components of the curvature reduce to
\begin{eqnarray}  
\widehat{R}_{0101} &=& \widehat{R}_{0202}=- \omega^2 
                          \label{rieh1st} \:, \\  
\widehat{R}_{1212} &=& m^2 - 3\,\omega^2 \label{rieh2nd} \:, \\
\widehat{R}_{3434} &=& k \label{riehlast} \:.
\end{eqnarray}
Following Cartan's method for the local equivalence,
we calculate the f\/irst covariant derivative of
the Riemann tensor. One obtains the following non-null
covariant derivatives of the curvature:
\begin{equation} \label{drieh}
\widehat{R}_{0112;1} = \widehat{R}_{0212;2}= 
                    \omega\, (m^2  - 4\,\omega^2) \:. 
\end{equation}
Clearly, regardless of the value of the constant $k\,$, the 
f\/irst covariant derivative of the curvature is algebraically 
expressible in terms of the Riemann tensor. Moreover, the number 
of functionally independent functions of the coordinates 
$x^\mu$ among the components of the curvature and its 
f\/irst covariant derivative is zero ($t_0=t_1=0$).
As far as the dimension of the residual isotropy group is 
concerned we distinguish three dif\/ferent classes of locally 
homogeneous 5D generalized G\"odel-type curved manifolds, 
according to the relevant parameters $m^2$, $\omega$ and 
$k\,$, namely~\cite{foot1}         
\begin{enumerate}
\item[1.] $\mbox{dim}\,(H_0) = \mbox{dim}\, (H_1)= 2\,$ when
  \begin{enumerate}
  \item[a)]
  $\,\omega \not=0\,$,  any real $k\,$,  $\,m^2 \not=4\,\omega^2\,$ ;  
  \item[b)]
  $\,\omega =0\,$,  $k\not= 0\,$, $\,m^2 \not=0\,$ ;
  \end{enumerate}  
\item[2.] $\,\mbox{dim}\,(H_0) = \mbox{dim}\, (H_1)= 4\,$ when
  \begin{enumerate}
  \item[a)]
  $\,\omega \not=0\,$, any real $k\,$,  $\,m^2=4\,\omega^2\,$ ;
  \item[b)]
  $\,\omega =0\,$,  $k=0\,$,  $\,m^2 \not=0\,$ ;
  \item[c)]
  $\,\omega =0\,$,  $k\not=0\,$, $\,m^2=0\,$ ;
  \end{enumerate}
\item[3.] $\,\mbox{dim}\,(H_0) = \mbox{dim}\, (H_1)= 10\,$ 
when $\omega = k = m^2 = 0\,$.
\end{enumerate}

Thus, from eqs.\ (\ref{gdim}) and (\ref{ddim}) one f\/inds that 
the locally homogeneous 5D generalized G\"odel-type 
manifolds admit a (local) $G_r$, with either $r =7$, $r=9\,$,
or $r=15$ acting on an orbit of dimension $d = 5$, that is on 
the manifold ${\cal M}_5$.

The above results can be collected together in the following 
theorems:

\vspace{2mm} 
\begin{theorem} \label{TheoHom}
The necessary and sufficient conditions for a five-dimensional 
generalized G\"odel-type manifold to be locally homogeneous 
are those given by  equations (\ref{cond1})~--~(\ref{cond3}).
\end{theorem}
\begin{theorem} \label{EquivTheo}
\begin{sloppypar}
The five-dimensional homogeneous generalized G\"o\-del-type 
manifolds are locally characterized by three independent real
parameters $\omega$, $k$ and $m^2\,$: identical triads 
($\omega, k,\, m^2$) specify locally equivalent manifolds.
\end{sloppypar}
\end{theorem}                      
\begin{theorem} \label{GroupTheo}
The five-dimensional locally homogeneous generalized G\"o\-del-type 
manifolds admit group of isometry $G_r$ with 
\begin{enumerate}
\item[(i)]
$r=7\:$  if either of the above conditions (1.a) and (1.b) is fulfilled;
\item[(ii)]
$r=9\:$  if one of the above set of conditions (2.a), (2.b) and (2.c) is 
fulfilled;
\item[(iii)]
$r=15\;$ if the above condition (3) is satisfied.
\end{enumerate}
\end{theorem}

We shall now focus our attention on the irreducible set of 
isometrically nonequivalent homogeneous generalized 
G\"odel-type metrics. These nonequivalent classes of
metrics can be obtained by a similar procedure to that used 
by Rebou\c{c}as and Tiomno~\cite{Reboucas83}, namely by 
integrating equations~(\ref{cond1})~--~(\ref{cond3}),
and eliminating through coordinate transformations the 
non-essential integration constants taking into account 
the relevant parameters according to the above 
theorem~\ref{EquivTheo}. 
For the sake of brevity, however, we shall only present 
the irreducible classes without going into details of 
calculations. It turns out that one ought to distinguish 
six classes of metrics according to:

{\bf Class I} : $\,m^2 > 0\,$, any real $k\,$, $\,\omega \not=0 $.
The line element for this class of homogeneous generalized 
G\"odel-type manifolds can always be brought [in cylindrical 
coordinates $(r, \phi, z)$] into the form 
\begin{equation} \label{ds2c}
d\hat{s}^{2}=[\,dt+H(r)\, d\phi\,]^{2} -D^{2}(r)\, d\phi^{2} -dr^{2} 
                 -F^2(u)\,dz^{2} - du^2
\end{equation}
with the metric functions given by
\begin{eqnarray} 
H(r) &=&\frac{2\,\omega}{m^{2}}\: [1 - \cosh\,(mr)]\;, \label{Hh} \\
D(r) &=& m^{-1}\, \sinh\,(mr) \label{Dh} \;,
\end{eqnarray}
\begin{eqnarray}     \label{Ffun}
F\,(u) = \left\{ \begin{array}
{l@{\qquad \mbox{if} \qquad}l}
\alpha^{-1}\, \sin\, (\alpha\, u )   & k = - \alpha^2 < 0 \;, \\
u         & k  =  0 \;, \\
\alpha^{-1}\, \sinh\, (\alpha\,u)   & k = \alpha^2 > 0 \;.
\end{array} \right.
\end{eqnarray}
According to theorem~\ref{GroupTheo} the possible isometry groups 
for this class are either $G_7$ (for $m^2 \not= 4\,\omega^2$) or
$G_9$ (when $m^2 = 4\,\omega^2$), irrespective of the value 
of $k\,$.

{\bf Class II} : $\,m^2 = 0\,$, any real $k\,$, $\,\omega \not=0 $. 
The line element for this class can be brought into the form~(\ref{ds2c}), 
with the metric function $F(u)$ given by~(\ref{Ffun}), but now the
functions $H(r)$ and $D(r)$ are given by
\begin{equation} \label{DHsr}
H(r) = - \,\omega\, r^{2}  \: \qquad \mbox{and} \: \qquad D(r) = r \:.
\end{equation}
For this class from theorem~\ref{GroupTheo} there is a group $G_7$ of 
isometries, regardless of the value of $k$.

{\bf Class III} : $\,m^{2} \equiv - \mu^{2} < 0\,$, any real $k\,$, 
                $\,\omega \not=0 $. 
Similarly for this class the line element reduces to ~(\ref{ds2c}) with
$F(u)$ given by (\ref{Ffun}) and
\begin{eqnarray}    
H(r) &=& \frac{2\,\omega}{\mu^{2}} \:[\cos\,(\mu r) - 1 ] \;, \label{Hc} \\ 
D(r) &=& \mu^{-1}\, \sin\,(\mu r)\;. \label{Dc}  
\end{eqnarray}
{}From theorem~\ref{GroupTheo}, regardless the value of $k$ for this class 
there is a group $G_7$ of isometries.

{\bf Class IV} : $\,m^{2} \not= 0\,$, any real $k\,$, and $\,\omega = 0$. 
We shall refer to this class as degenerated G\"odel-type manifolds,
since the cross term in the line element, related to the
rotation $\omega$ in 4D G\"odel model, vanishes. 
By a trivial coordinate transformation one can make $H = 0$ 
with $D(r)$ given, respectively, by (\ref{Dh}) 
or (\ref{Dc}) depending on whether $m^2>0$ or
$m^{2} \equiv - \mu^{2} < 0$. The function $F(u)$ depends on 
the sign of $k$ and is again given by (\ref{Ffun}).
For this class according to theorem~\ref{GroupTheo} one
may have either a $G_7$ for $k\not=0$, or a $G_9$ for
$k=0\,$.

{\bf Class V} : $\,m^{2} = 0\,$, $k \not=0 \,$, and $\,\omega = 0$.
By a trivial coordinate transformation one can make $H=0\,$, $D = r\,$
and $F(u) = \alpha^{-1}\,\sin\, (\alpha\,u)\,$ or 
$\,F(u) = \alpha^{-1}\sinh\, (\alpha\,u) \,$ depending on whether $\,k<0\,$ 
or $\,k>0\,$, respectively. From theorem~\ref{GroupTheo} there is 
a group $G_9$ of isometries.

{\bf Class VI} : $\,m^{2} = 0\,$,  $k=0\,$, and $\,\omega = 0$.
{}From (\ref{rieh1st})~--~(\ref{riehlast}) this corresponds to
the 5D f\/lat manifold. Therefore, one can make
$H=0\,$, $D(r)= r\,$ and $F(u) = u \,$.
Theorem~\ref{GroupTheo} ensures that there is a group $G_{15}$
of isometries.

\vspace{3mm}
\section{Killing Vector Fields}
\label{Killing}
\setcounter{equation}{0}

In this section we shall present  the inf\/initesimal
generators of isometries of the 5D homogeneous generalized 
G\"odel-type manifolds, whose line element (\ref{ds2c}) can 
be brought into the Lorentzian form (\ref{ds2f}) with
$\widehat{\Theta}^A$ given by
\begin{equation} \label{lorpen1}
\widehat{\Theta}^{0} = dt + H(r)\,d\phi\:, \: \quad
\widehat{\Theta}^{1} = dr\:, \: \quad
\widehat{\Theta}^{2} = D(r)\,d\phi\:, \:\quad
\widehat{\Theta}^{3} = F(u)\,dz \:, \: \quad
\widehat{\Theta}^{4} = du \:,
\end{equation}
where the functions $H(r)\,$, $D(r)$ and $F(u)$ depend
upon the essential parameters $m^2\,$, $k\,$ and $\omega\,$
according to the above classes of locally homogeneous 
manifolds.

Denoting the coordinate components of a generic Killing vector f\/ield 
$\widehat{K}$ by $\widehat{K}^{u} \equiv (Q, R, S,\bar{Z},U)$,
where $Q, R, S, \bar{Z} $ and $U$ are functions of all 
coordinates $t,r,\phi,z$, $u$, then the f\/ifteen Killing equations 
\begin{equation} \label{killeqs}
\widehat{K}_{(A;B)} \equiv \widehat{K}_{A;B} + \widehat{K}_{B;A} = 0 
\end{equation}
can be written in the Lorentz frame (\ref{ds2f})~--~(\ref{lorpen1}) 
as
\begin{eqnarray}  
&T_t  =  0 \:, \qquad T_{u} - U_t  =  0    \label{um} \:, \\
&R_r  =  0 \:, \qquad U_r   + R_{u}  =  0    \label{dois} \:, \\
&U_{u}  =  0   \:,   \label{tres} \\
&D\,(T_r - R_t) - H_r P  =  0  \label{quatro} \:,          \\
& D P_{u} + U_{\phi} - H U_t  =  0 \label{cinco} \:,        \\
&T_{\phi} + H_r R - D P_t  =  0 \label{seis}  \:,        \\
&R_{\phi} - H R_t - D_r P + D P_r  =  0  \label{sete} \:, \\
&P_{\phi} - H P_t + D_r R  =  0  \label{oito}        \:, \\ 
& T_z - F\,Z_t  =  0 \label{nove}   \:, \\
& F\, Z_r +  R_z  =  0  \label{dez}    \:, \\
& Z_z + U\, F_u =  0 \label{onze}   \:, \\
&U_z + F\, Z_{u} - Z\, F_u  =  0   \label{doze}  \:, \\
& D P_z + F\,( Z_{\phi} - H Z_t)  =  0  \label{treze}  \:,
\end{eqnarray}
where the subscripts denote partial derivatives, and where we 
have made
\begin{equation}  \label{pencor}
T \equiv H\,S + Q, \qquad P \equiv D\,S,  \qquad \mbox{and} \qquad 
Z \equiv F\, \bar{Z} 
\end{equation}
to make easier the comparison and the use of the results obtained 
in~\cite{Teixeira85}.
To this end we note that with the changes $u \rightarrow z$ and 
$U \rightarrow Z$ the above equations (\ref{um})~--~(\ref{oito}) 
are formally identical to the Killing equations (4) to (11) 
of~\cite{Teixeira85}. However, in the  equations%
~(\ref{um})~--~(\ref{oito}) the functions $T, R, P, U$ 
depend additionally on the f\/ifth coordinate $u$. Taking into 
account this similitude,  the integration of the 
Killing equations (\ref{um})~--~(\ref{treze}) can be obtained 
in two steps as follows. First, by analogy with (4) to (11) 
of Ref.~\cite{Teixeira85} one integrates (\ref{um})~--~(\ref{oito}), 
but at this step instead of the integration constants one has 
integration functions of the f\/ifth coordinate $u$. 
Second, one uses the remaining eqs.\ (\ref{nove})~--~(\ref{treze})
to achieve explicit forms for these integration functions and
to obtain the last component $U$ of the generic Killing vector $K$.

We have used the above two-steps procedure to integrate the 
Killing equations (\ref{um})~--~(\ref{treze}) for all class
of homogeneous generalized G\"odel-type manifolds. 
However, for the sake of brevity, we shall only present the 
Killing vector f\/ields and the corresponding Lie algebras without 
going into details of calculations, which can be verif\/ied 
by using, for example, the computer algebra program {\sc killnf},
written in {\sc classi} by {\AA}man~\cite{Aman87}. 

{\bf Class I} : $\,m^2 > 0\,$, any real $k\,$,  $\,\omega \not=0 $. 
In the integration of the Killing equation for this general class 
one is led to distinguish two dif\/ferent subclasses of solutions 
depending on whether $m^2 \not= 4 \,\omega^2$ or $m^2 =4 \,\omega^2$. 
We shall refer to these subclasses as classes Ia and Ib, respectively. 

{\bf Class Ia} : $\,m^2 >0\,$, any real $k\,$, $\,m^2\not=4\,\omega^2$. 
In the coordinate basis in which as (\ref{ds2c}) is given,
a set of linearly independent Killing vector f\/ields 
$K_N$ ($N$ is an enumerating index) is given by
\begin{eqnarray}
K_1 &=&\partial_t \:,  \qquad  \quad
K_2  = \frac{2\,\omega}{m}\, \,\partial_t - m \,\partial_{\phi} \:,
                         \label{KIa1} \\
K_3 &=& -\,\frac{H}{D}\, \sin\phi\, \,\partial_t +\cos\phi\, \,\partial_r
        -\,\frac{D_r}{D}\, \sin\phi\, \,\partial_{\phi} \:, 
                          \label{KIa2} \\
K_4 &=& -\,\frac{H}{D}\, \cos\phi\, \,\partial_t -\sin\phi\, \,\partial_r
        -\,\frac{D_r}{D}\, \cos\phi\, \,\partial_{\phi} \:, 
                           \label{KIa3} \\
K_5 &=& \sin z\, \,\partial_u +\frac{F_u}{F}\,\cos z\, \,\partial_z \:, 
                         \label{KIa4} \\
K_6 &=& \cos z\, \,\partial_u - \frac{F_u}{F}\,\sin z\, \,\partial_{z} \:, 
                         \label{KIa5} \\
K_7 &=& \partial_{z} \;. \label{KIa6}
\end{eqnarray}

The Lie algebra has the following nonvanishing commutators:
\begin{eqnarray}
& \left[ K_2, K_3 \right] = - m\, K_4  \:, \qquad
\left[ K_2, K_4 \right] =  m\, K_3   \:, \qquad
\left[ K_3, K_4 \right] = m\, K_2  \:, \\
&\left[ K_5, K_6 \right]    =  -\, k \, K_7 \:, \qquad
\left[ K_5, K_7 \right]  =  - K_6   \:, \qquad 
\left[ K_6, K_7 \right]  =  K_5   \:.
\end{eqnarray} 
Therefore the corresponding algebra is 
${\cal L}_{Ia} = {\cal L}_k \oplus \tau \oplus so\,(2,1)$.
Here and in what follows the symbols $\oplus\,$
and $\,\sds$ denote and direct and semi-direct sum of sub-algebras, 
and the sub-algebra ${\cal L}_k$ is $so\,(3)$ for $k < 0 \,$,
$so\,(2,1)$ for $k > 0 \,$, and $t^2 \,\sds \, so\,(2)\,$ for $k=0\,$.
For the present class  ${\cal L}_k$ is generated by
$K_5, K_6\,$ and  $K_7$, the symbol $\tau$ is associated to the time 
translation $K_1$, and f\/inally the inf\/initesimal generators of 
sub-algebra $so\,(2,1)$ are $K_2,\, K_3\,$ and $K_4$.

{\bf Class Ib} : $\,m^2 =4\, \omega^2\,$, any real $k\,$, $\,\omega\not=0$. 
For this class the Killing vector f\/ields are
\begin{eqnarray}
K_1 &=&\partial_t \:, \quad \qquad 
K_2 = \partial_t - m \,\partial_{\phi} \:,  
           \label{KIb1} \\
K_3 &=& -\,\frac{H}{D}\, \sin\phi\, \,\partial_t +\cos\phi\, \,\partial_r
     -\,\frac{D_r}{D}\, \sin\phi\, \,\partial_{\phi} \:, \label{KIb2} \\
K_4 &=& -\,\frac{H}{D}\, \cos\phi\, \,\partial_t -\sin\phi\, \,\partial_r
        -\,\frac{D_r}{D}\, \cos\phi\, \,\partial_{\phi} \:, \label{KIb3} \\
K_5 &=&-\,\frac{H}{D}\,\cos(mt+\phi)\,\,\partial_t 
          +\sin(mt+\phi)\,\,\partial_r
    +\,\frac{1}{D}\, \cos(mt+\phi)\,\,\partial_{\phi} \:, \label{KIb4} \\
K_6 &=&-\,\frac{H}{D}\,\sin(mt+\phi)\,\,\partial_t 
            -\cos(mt+\phi)\,\,\partial_r
    +\,\frac{1}{D}\, \sin(mt+\phi)\,\,\partial_{\phi} \:, \label{KIb5} \\
K_7 &=& \sin z\, \,\partial_u +\frac{F_u}{F}\,\cos z\, \,\partial_z \:, 
                         \label{KIb6} \\ 
K_8 &=& \cos z\, \,\partial_u - \frac{F_u}{F}\,\sin z\, \,\partial_{z} \:, 
                         \label{KIb7} \\
K_9 &=& \partial_{z} \:, \label{KIb8} 
\end{eqnarray}
whose Lie algebra is given by
\begin{eqnarray}
&\left[ K_1, K_5 \right] = -m\, K_6 \:, \qquad
\left[ K_1, K_6 \right] = m\, K_5 \:, \qquad
\left[ K_2, K_3 \right] = - m\, K_4   \:, \\
&\left[ K_2, K_4 \right] = m\, K_3  \:, \qquad 
\left[ K_3, K_4 \right]    =  m\, K_2 \:, \qquad 
\left[ K_5, K_6 \right]  = m\,  K_1  \:, \\
&\left[ K_7, K_8 \right]    =  -\, k \, K_9 \:, \qquad
\left[ K_7, K_9 \right]  =  - K_8   \:, \qquad 
\left[ K_8, K_9 \right]  =  K_7   \:. 
\end{eqnarray} 
So, the corresponding algebra for this case is 
${\cal L}_{Ib} = {\cal L}_k \oplus so\,(2,1) \oplus so\,(2,1)$.
As in the previous class the sub-algebra ${\cal L}_k$ depends on
the sign of $k$, and here is generated by $K_7,K_8$ and $K_9$.
The two sub-algebras $so\,(2,1)$ are generated by the Killing vector
f\/ields $K_1, K_5, K_6$ and  $K_2, K_3, K_4$.

{\bf Class II} : $\,m^2 = 0\,$, any real $k\,$, $\,\omega \not=0$. 
For this class the Killing vector f\/ields turns out to be the 
following:
\begin{eqnarray}
K_1 &=&\partial_t \:,  \quad \qquad
K_2 = \partial_{\phi} \:, \label{KII1} \\
K_3 &=& -\,\omega\,r\, \sin\phi\, \,\partial_t -\cos\phi\, \,\partial_r
    +\,\frac{1}{r}\, \sin\phi\, \,\partial_{\phi} \:, \label{KII2} \\
K_4 &=& -\,\omega\,r\, \cos\phi\, \,\partial_t +\sin\phi\, \,\partial_r
        +\,\frac{1}{r}\, \cos\phi\, \,\partial_{\phi} \:, \label{KII3} \\
K_5 &=& \sin z\, \,\partial_u +\frac{F_u}{F}\,\cos z\, \,\partial_z \:, 
                         \label{KII4} \\
K_6 &=& \cos z\, \,\partial_u - \frac{F_u}{F}\,\sin z\, \,\partial_{z} \:, 
                         \label{KII5} \\
K_7 &=& \partial_{z} \:. \label{KII6}
\end{eqnarray}
The Lie algebra has the following nonvanishing commutators:
\begin{eqnarray}
&\left[ K_2, K_3 \right] =  K_4   \:, \quad
\left[ K_2, K_4 \right] = - K_3  \:, \quad 
\left[ K_3, K_4 \right]    =  2\, \omega\,  K_1 \:, \\   
&\left[ K_5, K_6 \right]    =  -\, k \, K_7 \:, \qquad
\left[ K_5, K_7 \right]  =  - K_6   \:, \qquad 
\left[ K_6, K_7 \right]  =  K_5   \:. 
\end{eqnarray} 
Therefore, the corresponding algebra for this case is 
${\cal L}_{II} = {\cal L}_k \oplus {\cal L}_4$.
The sub-algebra ${\cal L}_4$ is generated by $K_1, K_2, K_3$ 
and $K_4$. This algebra ${\cal L}_4$ is soluble and 
does not contain abelian 3D sub-algebras; it is classif\/ied
as type $III$ with $q=0$ by Petrov~\cite{Petrov69}.  
The sub-algebra ${\cal L}_k$ is the same of the previous
classes and is generated by $K_5, K_6$ and $K_7$.

{\bf Class III} : $\,m^{2} \equiv - \mu^{2} < 0\,$, any real $k\,$,
                  $\,\omega \not=0 $. 
For this class the set of linearly independent Killling vector f\/ields
we have found is given by
\begin{eqnarray}
K_1 &=&\partial_t \:,  \quad \: 
K_2 = \frac{2\,\omega}{\mu} \, \partial_t 
             + \mu\, \partial_{\phi} \:, \label{KIII1} \\
K_3 &=& -\,\frac{H}{D}\, \sin\phi\, \,\partial_t 
            +\cos\phi\, \,\partial_r
   -\,\frac{D_r}{D}\, \sin\phi\, \,\partial_{\phi} \:, \label{KIII2} \\
K_4 &=& -\,\frac{H}{D}\, \cos\phi\, \,\partial_t 
            -\sin\phi\, \,\partial_r
   -\,\frac{D_r}{D}\, \cos\phi\, \,\partial_{\phi} \:, \label{KIII3} \\
K_5 &=& \sin z\, \,\partial_u +\frac{F_u}{F}\,\cos z\, \,\partial_z \:, 
                         \label{KIII4} \\
K_6 &=& \cos z\, \,\partial_u - \frac{F_u}{F}\,\sin z\, \,\partial_{z} \:, 
                         \label{KIII5} \\
K_7 &=& \partial_{z} \:. \label{KIII6}
\end{eqnarray}
The Lie algebra has the following nonvanishing commutators:
\begin{eqnarray}
&\left[ K_2, K_3 \right] =  \mu \, K_4   \:, \qquad
\left[ K_2, K_4 \right] = -\mu \, K_3  \:, \qquad 
\left[ K_3, K_4 \right]    =  \mu \, K_2 \:, \\   
&\left[ K_5, K_6 \right]    =  -\, k \, K_7 \:, \qquad
\left[ K_5, K_7 \right]  =  - K_6   \:, \qquad 
\left[ K_6, K_7 \right]  =  K_5   \:. 
\end{eqnarray} 
Thus, the corresponding algebra for this case is 
${\cal L}_{III} = {\cal L}_k \oplus \tau \oplus so\,(3)$.
Here $\tau$ is associated to the Killing vector f\/ield $K_1$,
whereas to the sub-algebra $so\,(3)$ correspond $K_2$,
$K_3$ and $K_4$. Again ${\cal L}_k$ is generated by
$K_5, K_6$ and $K_7$.

{\bf Class IV} : $\,m^2 \not= 0\,$, any real $k\,$,  $\,\omega=0 $. 
In the integration of the Killing equation for this general class 
one is led to distinguish two dif\/ferent subclasses according to 
$k\not=0$ or $k=0$. We shall denote these subclasses 
as classes IVa and IVb, respectively. 

{\bf Class IVa} : $\,m^{2} \not= 0\,$, $k\not=0$, $\,\omega = 0$. 
This class corresponds to the so-called degenerated 
G\"odel-type manifolds. One obtains for this class the 
following Killing vector f\/ields:
\begin{eqnarray}
K_1 &=&\partial_t \:,  \qquad \quad K_2 = \partial_\phi \:, 
                   \label{KIVa1} \\
K_3 &=& \cos\phi\, \,\partial_r-
   \,\frac{D_r}{D}\, \sin\phi\, \,\partial_{\phi} \:, 
                   \label{KIVa2} \\
K_4 &=& -\sin\phi\, \,\partial_r
  -\,\frac{D_r}{D}\, \cos\phi\, \,\partial_{\phi} \:, 
                    \label{KIVa3}  \\
K_5 &=& \sin z\, \,\partial_u +\frac{F_u}{F}\,\cos z\, \,\partial_z \:, 
                         \label{KIVa4} \\
K_6 &=& \cos z\, \,\partial_u - \frac{F_u}{F}\,\sin z\, \,\partial_{z} \:, 
                         \label{KIVa5} \\
K_7 &=& \partial_{z} \:, \label{KIVa6}
\end{eqnarray}
where $D(r) = (1/m)\, \sinh mr$ for $m^2>0\,$,  or 
$\,D(r) = (1/\mu)\, \sin \mu r$ for $m^2 \equiv - \mu^2 < 0 $,
and the function $F(u)$ for $k\not=0$ is given by~(\ref{Ffun}).
The Lie algebra has the following nonvanishing commutators:
\begin{eqnarray}
&\left[ K_2, K_3 \right] =  K_4   \:, \qquad
\left[ K_2, K_4 \right] = - K_3  \:, \qquad 
\left[ K_3, K_4 \right]    = - m^2 \, K_2 \:, \\   
&\left[ K_5, K_6 \right]    =  -\, k \, K_7 \:, \qquad
\left[ K_5, K_7 \right]  =  - K_6   \:, \qquad 
\left[ K_6, K_7 \right]  =  K_5   \:,
\end{eqnarray} 
where one should substitute $-m^2$ by $\mu^2$ if $m^2 < 0 $. So, the 
corresponding Lie algebra is ${\cal L}_{IVa}={\cal L}_k \oplus 
\tau \oplus {\cal L}_m$, where ${\cal L}_m$ is $so\,(2,1)$ for $m^2>0$,
and $so\,(3)$ for $m^2 = - \mu^2 <0$. The sub-algebra ${\cal L}_k$
(generated by $K_5, K_6$ and $K_7$) is $so\,(3)$  for $k < 0\,$,
and $so\,(2,1)$ for $k > 0 \,$.  Again $\tau$ is associated to the 
Killing vector f\/ield $K_1$.

{\bf Class IVb} : $\,m^{2} \not= 0\,$, $k=0$, $\,\omega = 0$. 
We shall refer to this class as doubly-degenerated 
G\"odel-type manifolds. One obtains for this class the 
following Killing vector f\/ields:
\begin{eqnarray}
K_1 &=&\partial_t \:,  \qquad \quad K_2 = \partial_\phi \:, 
                   \label{KIVb1} \\
K_3 &=& \cos\phi\, \,\partial_r-
   \,\frac{D_r}{D}\, \sin\phi\, \,\partial_{\phi} \:, 
                   \label{KIVb2} \\
K_4 &=& -\sin\phi\, \,\partial_r
  -\,\frac{D_r}{D}\, \cos\phi\, \,\partial_{\phi} \:, 
                    \label{KIVb3}  \\
K_5 &=& \sin z\, \,\partial_u +\frac{1}{u}\,\cos z\, \,\partial_z \:, 
                         \label{KIVb4} \\
K_6 &=& \cos z\, \,\partial_u - \frac{1}{u}\,\sin z\, \,\partial_{z} \:, 
                         \label{KIVb5} \\
K_7 &=& \partial_{z}  \:, \label{KIVb6} \\
K_8 &=& u\,\sin z\, \,\partial_t + t \, \sin z\, \,\partial_u  
            + \frac{1}{u}\,\,t\, \cos z \, \,\partial_z \:, 
                           \label{KIVb7} \\
K_9 &=& u\,\cos z\, \,\partial_t + t\,\cos z\, \,\partial_{u}  
            - \frac{1}{u}\,\,t\, \sin z\, \,\partial_z \:, 
                           \label{KIVb8} 
\end{eqnarray}
where  again $D(r) = (1/m)\, \sinh mr$ for $m^2>0\,$,  or 
$\,D(r) = (1/\mu)\, \sin \mu r$ for $m^2 \equiv - \mu^2 < 0 $.

The Lie algebra has the following nonvanishing commutators:
\begin{eqnarray}
&\left[ K_2, K_3 \right] =  K_4   \:, \qquad
\left[ K_2, K_4 \right] = - K_3  \:, \qquad 
\left[ K_3, K_4 \right]    = - m^2 \, K_2 \:, \\   
&\left[ K_5, K_7 \right]  =  - K_6   \:, \qquad 
\left[ K_6, K_7 \right]  =  K_5   \:, \qquad 
\left[ K_1, K_8 \right] =  K_5  \:,  \\
&\left[ K_1, K_9 \right]  =  K_6  \:,  \qquad
\left[ K_5, K_8 \right]  =  K_1 \:,    \qquad
\left[ K_6, K_9 \right] =  K_1   \:,   \\
&\left[ K_7, K_8 \right]  =  K_9  \:,   \qquad
\left[ K_7, K_9 \right]  = - K_8  \:,   \qquad
\left[ K_8, K_9 \right]  = - K_7  \:,  
\end{eqnarray} 
where one should substitute $-m^2$ by $\mu^2$ if $m^2 < 0 $.
So, the corresponding Lie algebra is 
${\cal L}_{IVb} = t^3\, \sds\, so\,(2,1) \oplus {\cal L}_m$, where 
${\cal L}_m$ is generated by $K_2, K_3, K_4$, and is either
$so\,(2,1)$ or $so\,(3)$ depending on whether $m^2>0$ or 
$m^2 = -\mu^2 <0$. The sub-algebra $t^3\,\sds\, so\,(2,1)\,$ 
is generated  by $K_1, K_5, K_6, K_7, K_8, K_9$.

{\bf Class V} : $\,m^{2} = 0\,$, $k\not=0$, $\,\omega = 0$. 
A set of linearly independent Killing vector f\/ield for this
class is
\begin{eqnarray}
K_1 &=&\partial_t \:,  \qquad \quad K_2 = \partial_\phi \:, 
                   \label{KV1} \\
K_3 &=& \cos\phi\, \,\partial_r-
   \,\frac{1}{r}\, \sin\phi\, \,\partial_{\phi} \:, 
                   \label{KV2} \\
K_4 &=& -\sin\phi\, \,\partial_r
  -\,\frac{1}{r}\, \cos\phi\, \,\partial_{\phi} \:, 
                    \label{KV3}  \\
K_5 &=& \sin z\, \,\partial_u +\frac{F_u}{F}\,\cos z\, \,\partial_z \:, 
                         \label{KV4} \\
K_6 &=& \cos z\, \,\partial_u - \frac{F_u}{F}\,\sin z\, \,\partial_{z} \:, 
                         \label{KV5} \\
K_7 &=& \partial_{z}  \:, \label{KV6} \\
K_8 &=& r\,\sin \phi\, \,\partial_t + t \, \sin\phi\, \,\partial_r  
            + \frac{1}{r}\,\,t\, \cos \phi \, \,\partial_\phi \:, 
                          \label{KV7} \\
K_9 &=& r\,\cos \phi\, \,\partial_t + t\,\cos \phi\, \,\partial_r  
            - \frac{1}{r}\,\,t\, \sin \phi\, \,\partial_\phi \:, 
                           \label{KV8} 
\end{eqnarray}
where $F(u)$ depends upon the sign of $k$ and is given by eq.~(\ref{Ffun}).

The Lie algebra has the following nonvanishing commutators:
\begin{eqnarray}
&\left[ K_2, K_3 \right] =  K_4   \:, \qquad
\left[ K_2, K_4 \right] = - K_3  \:, \qquad 
\left[ K_5, K_6 \right]    =  -\, k \, K_7 \:, \\
&\left[ K_5, K_7 \right]  =  - K_6   \:, \qquad 
\left[ K_6, K_7 \right]  =  K_5   \:, \qquad 
\left[ K_1, K_8 \right] = - K_4  \:,  \\
&\left[ K_1, K_9 \right]  =  K_3  \:,  \qquad
\left[ K_4, K_8 \right]  = - K_1 \:,    \qquad
\left[ K_3, K_9 \right] =  K_1   \:,   \\
&\left[ K_2, K_8 \right]  =  K_9  \:,   \qquad
\left[ K_2, K_9 \right]  = - K_8  \:,   \qquad
\left[ K_8, K_9 \right]  = - K_2  \:.  
\end{eqnarray} 
So, the corresponding Lie algebra is 
${\cal L}_{V} = t^3\, \sds\, so\,(2,1) \oplus {\cal L}_k$, where 
${\cal L}_k$ is generated by $K_5, K_6, K_7$, and is either
$so\,(2,1)$ or $so\,(3)$ depending on whether $k>0$ or 
$k<0$. The sub-algebra $t^3\,\sds\, so\,(2,1)\,$ 
is generated by $K_1, K_2, K_3, K_4, K_8, K_9$.

{\bf Class VI} : $\,m^{2} = 0\,$, $k= 0$, $\,\omega = 0$.
{}From (\ref{rieh1st})~--~(\ref{riehlast}) this case corresponds to
the 5D f\/lat manifold whose Lie algebra is 
${\cal L}_{VI} = so\,(4,1)\,$ since it clearly has the well 
known f\/ifteen Killing vector f\/ields, namely f\/ive translations, 
four spacetime rotations, and six space rotations.

It is worth noting that none of the above Lie algebras
is semi-simple, but some of their sub-algebras are.
Besides, most of the simple sub-algebras are noncompact.
The 3D sub-algebra $so\,(3)$ present in all classes
is compact, though.

The number of Killing vector f\/ields we have found for each
of the above six classes makes explicit that the 5D locally 
homogeneous generalized G\"odel-type manifolds admit a 
group of isometry $G_7$ when 
({\em 1a\/}):
$\: m^2 \not = 4\,\omega^2\,$, any real $k\,$, $\,\omega \not=0$,
or when
({\em 1b\/}):
$\,m^2 \not=0$, $k\not=0$, $\omega=0\,$.
Groups $G_9$ of isometry occur when
({\em 2a\/}):
$\:\,m^2 = 4\, \omega^2$, any real $k$, $\omega \not=0$, or
({\em 2b\/}):
$m^2 \not= 0$,  $k = 0$, $\omega =0$, or when
({\em 2c\/}):
$m^2 = 0$,  $k \not= 0$, $\omega =0$. 
Clearly when $\: m^2 = \omega = k = 0\,$ there is $G_{15}$. 
These possible groups are in agreement with theorem~\ref{GroupTheo} 
of the previous section. 
Actually the integration of the Killing equations constitutes a 
dif\/ferent way of deriving that theorem. Furthermore, these 
equations also show that the isotropy subgroup $H$ of $G_r$ 
is such that $\,\mbox{dim}\,(H) = 2\,$  when the above conditions
({\em 1a\/}) and ({\em 1b\/}) are satisf\/ied,  while the conditions
({\em 2a\/}), ({\em 2b\/}) and ({\em 2c\/}) lead to
$\,\mbox{dim}\,(H) = 4\,$, also in agreement with the 
previous section.  Clearly $\,\mbox{dim}\,(H) = 10\,$
when $m^2 = \omega = k = 0$.

\vspace{3mm}  
\section{Causal Anomalies and Final Remarks}
\label{Anom}
\setcounter{equation}{0}
In this section we shall initially be concerned with the problem of
causal anomalies in the generalized G\"odel-type manifolds.
Then we proceed by examining whether the IM gravity 
allows solutions of generalized G\"odel-type metrics~(\ref{ds2c}).
Finally, we conclude by addressing to the general question as to whether 
the IM gravity theory rules out the 4D noncausal G\"odel-type 
solutions to Einstein's equations of general relativity.

In the f\/irst three of the six classes of homogeneous generalized 
G\"odel-type manifolds we have discussed in Section~\ref{homoconds},
there are closed timelike curves. 
Indeed, the analysis made in a previous paper%
~\cite{ReboucasTeixeira98a} can be easily extended to
the generalized 5D G\"odel-type manifolds of the present
article. To this end, we write the line element~(\ref{ds2c}) 
in the form
\begin{equation} \label{ds2cx} 
ds^2=dt^2 + 2\,H(r)\, dt\,d\phi -dr^2 -G(r)\,d\phi^2 
                 - F^2(u)\,dz^2 - du^2\,,
\end{equation}
where $G(r)= D^2 - H^2$ and $(r,\phi, z)$ are cylindrical 
coordinates.
Now, the existence of closed timelike curves of the G\"odel-type depends
on the behavior of $G(r)$. Indeed, if  $G(r) < 0$ for a certain
range of $r$ ($r_1 < r < r_2$, say), G\"odel's circles%
~\cite{Calvao88} $u,t,z,r =const$ are closed timelike curves. 

Since one can always make $H=0$ for the generalized G\"odel-type 
manifolds of classes IV, V and VI, then $G(r) > 0$ for all $r>0$. Thus
there are no closed timelike G\"odel's circles in these classes of
manifolds.

On the other hand, following the above-outlined reasoning it easy to 
show (see~\cite{ReboucasTeixeira98a} for details) that for each of the
remaining three classes (Class I to Class III) one can always f\/ind a 
critical radius $r_c$ such that for all $r > r_c$ one has $G(r) < 0$, 
making clear that there are closed timelike curves in these 
families of homogeneuous generalized G\"odel-type manifolds. 
However, in what follows we shall show that these types of
noncausal {\em curved\/} manifolds are not permitted 
in the context of the induced matter theory.  

In the Lorentz frame $\widehat{\Theta}^A$ given by~(\ref{lorpen1})
the nonvanishing frame components of the Einstein tensor 
$\widehat{G}_{AB}=\widehat{R}_{AB}-\frac{1}{2}\,R\,\widehat{\eta}_{AB}$ 
are
\begin{eqnarray}  
\widehat{G}_{00} &=& - \,\frac{D''}{D} + \frac{3}{4}\,
            \left( \frac{H'}{D}\,\right)^2       
             - \frac{\ddot{F}}{F} \label{ein00} \;, \\
\widehat{G}_{02} &=&  \frac{1}{2} \, \left( \frac{H'}{D}\, \right)' \;, 
                                                  \label{ein02} \\  
\widehat{G}_{11} &=& \widehat {G}_{22}\; = \; \frac{1}{4} \, 
                \left( \frac{H'}{D}\, \right)^2
             + \frac{\ddot{F}}{F} \:, \label{ein11} \\
\widehat{G}_{33} &=& \widehat{G}_{44}\; = \; \frac{D''}{D} - \frac{1}{4}\,
            \left( \frac{H'}{D}\,\right)^2         \label{ein33} \;, 
\end{eqnarray}
where the prime and dot denote derivative with respect to $r$
and $u$, respectively.

The f\/ield equations~(\ref{5DfeqsG}) require that 
$\widehat{G}_{02}=0$, which in turn implies that
\begin{equation}   \label{G02}
 \frac{H'}{D} = \mbox{const} \equiv - 2\,\omega\;.
\end{equation}
Inserting~(\ref{G02}) into~(\ref{ein11}), (\ref{ein33}) and (\ref{ein00})
one easily f\/inds that the IM f\/ield equations are fulf\/illed
if and only if  the independent parameters $\omega$, $k$ and $m^2$ 
[ see eqs.~(\ref{cond1})~--~(\ref{cond2} ] vanish identically,
which leads to the only solution given by

\begin{equation} \label{sol}
H =  a \;,  \qquad 
D =  b\,r + c \;, \qquad \mbox{and}  \qquad
F = \beta \, u + \gamma, 
\end{equation}
where $a$, $b$, $c$, $\beta$, and  $\gamma$ are arbitrary real 
constants. However, these constants have no physical meaning, and 
can be taken to be $a = c = \gamma = 0$ and $b=\beta= 1$ by a 
suitable choice of coordinates. 
Indeed, if one performs the coordinate transformations 
\begin{eqnarray} 
t   &  = & \bar{t} - \frac{a}{b}\,\, \bar{\phi}\:, \qquad \qquad     
r      =  \bar{r} - \frac{c}{b}\:,   \label{tr}    \\
\phi & = & \frac{\bar{\phi}}{b}\,  \:, \; \quad   
z   = \frac{\bar{z}}{\beta}  \:,  \qquad  
u  =  \bar{u}- \frac{\gamma}{\beta}  \:, \label{ppz}
\end{eqnarray}
the line element~(\ref{ds2cx}) becomes
\begin{equation}  \label{ds2flat}
d\hat{s}^2=d\bar{t}^2 -d\bar{r}^2 -\bar{r}^2 \,d\bar{\phi}^2 -
                 d\bar{z}^2 -d\bar{u}^2\;,
\end{equation}
in which we obviously have  $\,G(\bar{r})= \bar{r}^2 >0\,$ for 
$\bar{r} \not=0$. 
The line element (\ref{ds2flat}) corresponds to a 
manifestly f\/lat 5D manifold, making it clear that the underlying 
manifold can be taken to be the simply connected Euclidean 
manifold $\R^5$, and therefore as $\,G(\bar{r})>0\,$
no closed timelike circles are permitted. 
Furthermore the above results clearly show that the IM theory 
does not admit any {\em curved\/} 5D G\"odel-type 
metric~(\ref{ds2c}) as solution to its f\/ield
equations~(\ref{5DfeqsG}).

However, in a recent work Mc Manus~\cite{McManus94} has shown 
that a one-parameter family of solutions of the  f\/ield 
equations~(\ref{5DfeqsG}) previously found by Ponce de Leon%
~\cite{Leon88} was in fact f\/lat in five dimensions.
And yet the corresponding 4D induced models were shown to be a 
perfect f\/luid family of Friedmann-Robertson-Walker curved models
(see Refs.~\cite{Wesson96a,Wesson92c,Coley95} and also%
~\cite{Abolghasem96}~--~\cite{Liu98}, where other Riemann-f\/lat 
solutions are also discussed). 

Therefore a question which naturally arises here is whether the above 
5D f\/lat metric, which is the only solution to the IM f\/ield 
equations, can similarly give rise to any 4D {\em curved\/} 
spacetime. However, from~(\ref{ds2flat}) one obviously
has that the corresponding 4D spacetime is nothing but
the Minkowski f\/lat space (this result can also
be derived by using a computer algebra package as, e.g.,  
{\sc clas\-si}~\cite{Aman87,MacCallumSkea94} to calculate
the 4D curvature tensor for $m^2=\omega=0\,$). 
In brief, the only solution of the IM f\/ield equations~(\ref{5DfeqsG}) 
of generalized G\"odel-type is  the 5D f\/lat space~(\ref{sol}), which
give rise only to the 4D Minkowski (f\/lat) spacetime,
whose topology can be taken to be the simply connected Euclidean 
$\R^5$, in which no closed timelike curves are permitted.

Although the above results can be looked upon as if the induced 
matter theory works as an ef\/fective therapy for the causal 
anomalies which arises when one starts from the specif\/ic 
generalized 5D G\"odel-type family of metrics~(\ref{ds2cx}), 
this does not ensure that the induced matter version
of general relativity is an ef\/f\/icient treatment for the causal
anomalies (solutions with closed timelike curves) in general 
relativity as it has been conjectured in~\cite{ReboucasTeixeira98b}. 
Actually, in a recent paper (which unfortunately has not been 
initially noticed by Rebou\c{c}as and Teixeira%
~\cite{ReboucasTeixeira98b}) Romero {\em et al.}~\cite{Romero96} 
(see also~\cite{Lidsey97}) have shown that the induced matter 5D 
scheme is indeed general enough to locally generate all solutions 
to 4D Einstein's f\/ield equations. This is ensured by a theorem
due to Campbell~\cite{Campbell26} which  states that any analytic 
$n$-dimensional Riemannian space can be locally embedded in a 
$(n+1)$-dimensional Ricci-f\/lat space. 
In our context this amounts to saying that there must exist 
a f\/ive-dimensional Ricci-f\/lat space which locally gives 
rise to the 4D G\"odel noncausal solution of Einstein's 
equations of general relativity. 
Thus, what still remains to be done regarding G\"odel-type
spaces is to f\/ind out this 5D Ricci-f\/lat 
space which gives rise (locally) to the 4D G\"odel-type spacetimes 
of general relativity.

To conclude it is worth stressing some features of the local
underlying embedding of the induced matter theory. Any 
Riemann-f\/lat manifold obviously is also Ricci-f\/lat. The 
reverse, however, does not necessarily holds, and one can have 
Ricci-f\/lat spaces which are not Riemann-f\/lat. 
For the generalized 5D G\"odel-type geometries we have discussed 
in this paper the condition for Ricci-f\/latness 
($\widehat{R}_{AB}=0$) necessarily leads to Riemann-f\/lat spaces.
Remarkably many solutions of the f\/ield equations~(\ref{5DfeqsG}) 
are indeed Riemann-f\/lat (see~\cite{Wesson96a,McManus94,Coley95}
and ~\cite{Leon88}~--~\cite{Liu98}). 
{}From a purely mathematical 5D point of view all Riemann-f\/lat 
spaces are locally equivalent (locally isometric).
However, from the viewpoint of the 5D induced matter gravity 
all the above-referred 5D Riemann-f\/lat solutions give rise 
to physically (and geometrically) distinct 4D spacetimes%
~\cite{Wesson96a,McManus94,Coley95}, ~\cite{Leon88}~--~\cite{Liu98}. 
On the other hand, in the light of the equivalence problem 
techniques we have discussed in Section~\ref{prereq}, these 
5D Riemann-f\/lat examples also show that all 5D Cartan 
scalars~(\ref{CartanScl}) can vanish identically, with or 
without the vanishing of the corresponding (induced) 4D 
Cartan scalars.

\vspace{3mm}

\section*{Acknowledgement} \label{acknowl}  
\setcounter{equation}{0}

The authors are grateful to the scientif\/ic agency CNPq for 
the grants under which this work was carried out.
M.J. Rebou\c{c}as thanks C. Romero, V.B. Bezerra and J.B. 
Fonseca-Neto for motivating and fruitful discussions.


\begin{thebibliography}{99}
\bibitem{foot1} The integration of the Killing equations for these
classes of generalized G\"odel-type manifolds, which will be
performed in the next section, shows that
the isotropy group indeed depends on these special relations 
between the essential parameters $\,\omega$, $k$ and $m^2\,$.
\end{thebibliography}
\end{document}